\documentclass[sigconf]{acmart}

\usepackage{tcolorbox}
\usepackage{amsmath}
\usepackage{algorithm}
\usepackage{adjustbox}
\usepackage{graphicx}
\usepackage{xspace}

\AtBeginDocument{%
  }

\copyrightyear{2024}
\acmYear{2024}
\setcopyright{acmlicensed}\acmConference[ASE '24]{39th IEEE/ACM International Conference on Automated Software Engineering }{October 27-November 1, 2024}{Sacramento, CA, USA}
\acmBooktitle{39th IEEE/ACM International Conference on Automated Software Engineering (ASE '24), October 27-November 1, 2024, Sacramento, CA, USA}
\acmDOI{10.1145/3691620.3695349}
\acmISBN{979-8-4007-1248-7/24/10}

\newcommand{\repair}{\hbox{\textsc{ContractTinker}}\xspace}

\newcommand{\obs}[1]{   \begin{tcolorbox}[colframe=black, colback=gray!10, boxrule=1pt, left=1pt,right=1pt,top=1pt,bottom=1pt]   \textit{#1} \end{tcolorbox} }

\newcommand{\eg}{{\textit{e.g.}}\xspace}

\begin{document}

\title{ContractTinker: LLM-Empowered Vulnerability Repair for Real-World Smart Contracts}


\author{Che Wang}
\email{chewang@stu.pku.edu.cn}
\affiliation{%
  \institution{School of Computer Science \\ Peking University}
  \city{Beijing}
  \country{China}
}

\author{Jiashuo Zhang}
\email{zhangjiashuo@pku.edu.cn}
\affiliation{%
  \institution{School of Computer Science \\ Peking University}
  \city{Beijing}
  \country{China}
}

\author{Jianbo Gao}
\authornote{Corresponding author.}
\email{gao@bjtu.edu.cn}
\affiliation{%
  \institution{Beijing Key Laboratory of Security and Privacy in Intelligent Transportation \\ Beijing Jiaotong University}
  \city{Beijing}
  \country{China}
}

\author{Libin Xia}
\email{lbxia@stu.pku.edu.cn}
\affiliation{%
  \institution{School of Computer Science \\ Peking University}
  \city{Beijing}
  \country{China}
}

\author{Zhi Guan}
\email{guan@pku.edu.cn}
\affiliation{%
  \institution{National Engineering Research Center For Software Engineering \\ Peking University}
  \city{Beijing}
  \country{China}
}

\author{Zhong Chen}
\email{zhongchen@pku.edu.cn}
\affiliation{%
  \institution{School of Computer Science \\ Peking University}
  \city{Beijing}
  \country{China}
}

\renewcommand{\shortauthors}{Che et al.}

\begin{abstract}
\vspace{-0.25em}
Smart contracts are susceptible to being exploited by attackers, especially when facing real-world vulnerabilities.
To mitigate this risk, developers often rely on third-party audit services to identify potential vulnerabilities before project deployment.
Nevertheless, repairing the identified vulnerabilities is still complex and labor-intensive, particularly for developers lacking security expertise. 
Moreover, existing pattern-based repair tools mostly fail to address real-world vulnerabilities due to their lack of high-level semantic understanding. 
To fill this gap, we propose \repair, a Large Language Models (LLMs)-empowered tool for real-world vulnerability repair. 
The key insight is our adoption of the Chain-of-Thought approach to break down the entire generation task into sub-tasks. 
Additionally, to reduce hallucination, we integrate program static analysis to guide the LLM. 
We evaluate \repair on 48 high-risk vulnerabilities. The experimental results show that among the patches generated by \repair, 23 (48\%) are valid patches that fix the vulnerabilities, while 10 (21\%) require only minor modifications. 
A video of \repair is available at \url{https://youtu.be/HWFVi-YHcPE}.
\vspace{-0.5em}

\end{abstract}



\maketitle

\vspace{-1em}
\section{Introduction}
Smart contracts are highly susceptible to various types of attacks due to their immutable nature and their close association with financial activities on blockchains.
To prevent smart contracts from being attacked, developers of real-world applications often rely on third-party security audit services, \eg, Code4Rena\cite{code4rena}, to identify security vulnerabilities before deploying the contracts.

However, even with the information reported by the third-party audit service, vulnerability repair is a complex and labor-intensive task, especially for developers who may not be familiar with security attacks.
Consequently, automated program repair (APR) has emerged as a highly popular research topic with immense practical value.
It simplifies developers' tasks by automatically generating patches for specific types of vulnerabilities.

Several APR tools~\cite{sguard,evmpatch} have been proposed to assist smart contract developers in implementing patches. These tools typically target vulnerabilities with clear low-level characteristics, such as \textit{re-entrancy} and \textit{integer overflow}, and generate patches based on predefined patterns. However, as shown in a recent empirical study~\cite{zhang}, real-world vulnerabilities are often high-level functional bugs that are closely related to the business logic of the contract. Due to a lack of understanding of these high-level semantics, existing pattern-based methods may fail to effectively address real-world vulnerabilities.

\begin{figure*}[!htbp]
    \centering
    \includegraphics[width=0.8\textwidth]{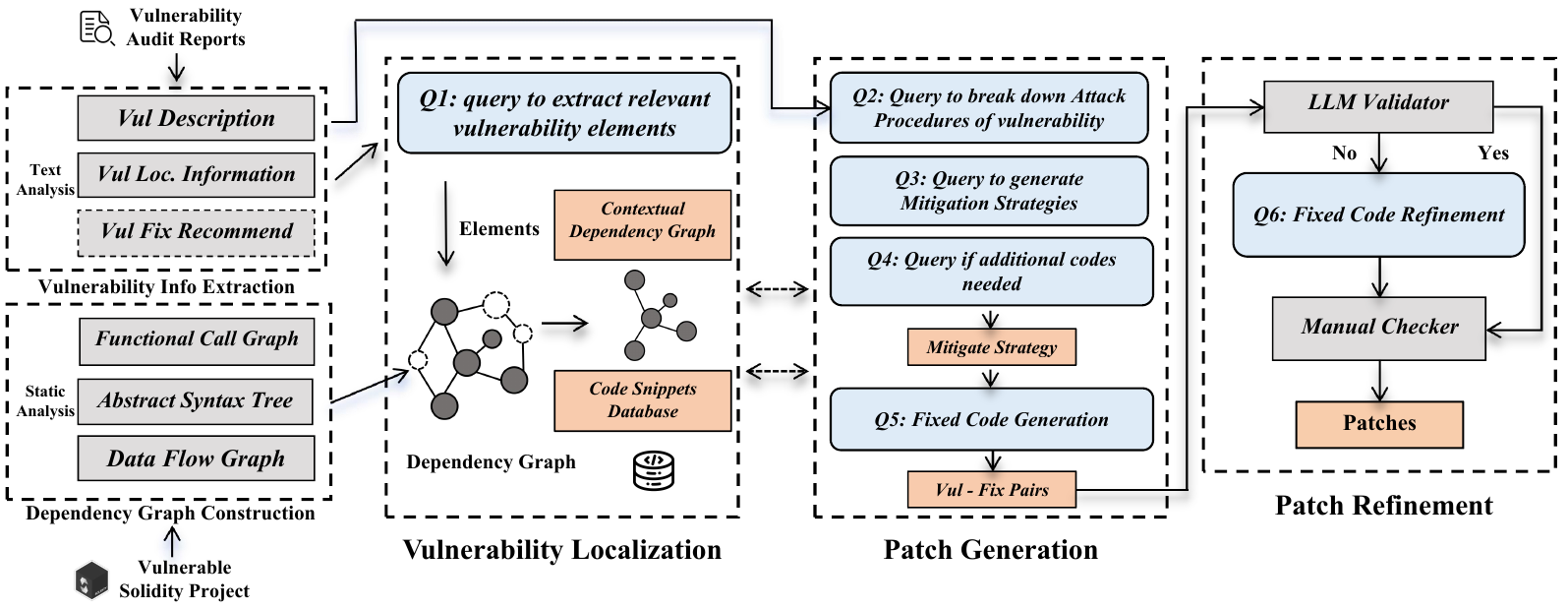}
\vspace{-2em}
    \caption{Workflow of \repair}
    \label{fig:framework}
\vspace{-1em}
\end{figure*}

To fill this gap, we developed \repair, a vulnerability repair tool for real-world smart contracts empowered by LLMs.
\textsc{ContractTinker} utilizes Large Language Models (LLMs) to understand high-level business logic of smart contracts, and extract structural information and vulnerability context from audit reports and smart contract code, respectively.
To mitigate the impact of LLM's hallucination, we break down the entire complex task through simulating the behavior of smart contract expert, and infer it step by step as presented in section~\ref{cot}. 
Additionally, in each step we integrate structural information and program static analysis results to guide the LLMs in generating fine-grained patches. We summarize our contributions as follows:

\setlength{\leftmargini}{0.6cm}
\begin{itemize}
    \item We proposed an LLMs-empowered tool, \repair, for real-world smart contract vulnerability repair. We employed the Chain-of-Thought (CoT) mechanism to guide the LLM in mapping vulnerability descriptions in audit reports to the semantics of smart contracts and generating corresponding patches for these vulnerabilities.
    \item We incorporated static analysis, including dependency analysis and program slicing, to help LLMs localizing the vulnerability in complex real-world smart contract projects and improve the efficiency of patch generation.
    \item We open-sourced our tool and dataset at \url{https://github.com/CheWang09/LLM4SMAPR}.
    The preliminary evaluation on 48 real-world functional vulnerabilities demonstrates the effectiveness of \repair. Of these, 23 (48\%) are valid patches that fix the vulnerabilities, while 10 (21\%) require only minor modifications. 
\end{itemize}
\vspace{-2mm}

\section{Approach} 
The workflow of \repair is illustrated in Figure~\ref{fig:framework}.
Firstly, users input the project and audit report. Then \repair performs two steps: It analyzes the project to extract an entire dependency graph and extracts the valid structural information from reports. 
Upon obtaining the necessary elements, the tool performs vulnerability localization to filter out useless elements, and builds a contextual dependency graph (CDG) as well as program slices. During patch generation, we adopted the "chain of thought" concept, breaking down the task to implement patch inference step by step. Finally, the tool employs another independent LLM to evaluate the generated patch code and refine it if it does not fix the vulnerability. The final output is validated patches.

\subsection{\textbf{Dependency Analysis}}
Since data and code are coupled in Solidity-based smart contracts, the interactions between functions and state variables, as well as among functions themselves, are crucial to understanding business logic errors. Therefore, to construct the dependency graph of the entire project, we defined three core elements included in dependency graph: function ($F$), state variables ($V$), and contracts ($C$). We define five types of edges in graph: $F-F$, $F-V$, $V-V$, $C-F$, $C-V$. Eventually, the dependency graph is defined as $DG = \{<V, E>| V\in \{F;V;C\},\ E\in \{[F,V]-[F,V];C-[F,V]\} \}$. 

When constructing a dependency graph, we mainly leverage three types of program analysis information: the Functional Call Graph, the Abstract Syntax Tree, and the Data Flow Graph. The Functional Call Graph is first utilized to build a base dependency graph of callee-caller between functions and contracts ($F-F$,$C-F$). Next, we extract the read-write relationship between functions/contracts and variables ($F-V$,$C-V$) that exist within the Abstract Syntax Tree. Finally, we extract data flow relationships among variables ($V-V$) from the Data Flow Graph to supply the dependency graph.
These program structural information are utilized in Section~\ref{val} and Section~\ref{cot} to locate vulnerabilities and generate patches, respectively.

\vspace{-2.5em}
\subsection{\textbf{Vulnerability Analysis \& Localization}} \label{val}
This section aims to locate vulnerability scope in project by combining extracted vulnerability information from text and dependency graph.
First, we extract the vulnerability description from the audit report. Afterwards, we extract program information included in the description, such as vulnerability locations. This process is performed from two perspectives: First, it extracts explicitly pointed out vulnerable functions. Secondly, if the vulnerable function is not explicitly pointed out, we apply entity recognition \cite{ner} on the vulnerability description via LLMs, as shown in Figure~\ref{fig:framework} Q1, to extract relevant contracts, functions, and state variables.

Based on the above two preprocessed data sources, we employ program slicing, as used in previous studies \cite{programslicing}, to localize the scope of vulnerable functions. We prioritize explicitly pointed-out functions as target functions to retrieve relevant elements in dependency graph. Then, we set recognized elements as seeds to retrieve relevant elements as supplement. 
At Last, we construct a contextual dependency graph (CDG), which is a pruned-graph of the dependency graph containing elements closely related to vulnerabilities. We create program slices consisting of all relevant code snippets of CDG.

\vspace{-1em}
\subsection{\textbf{Patch Generation}}\label{cot}
Generally, a single patch generation task involves several internal inference steps, such as vulnerability localization, analysis, and patch generation. 
However, it is complex for the LLM to perform this reasoning, which could result in hallucination. Therefore, we employ the Chain of Thoughts approach as suggested by previous studies \cite{CoT}. We simulated the behavior of smart contract experts and broke down patch generation into several reasoning steps as shown in Figure~\ref{fig:framework}: vulnerability attack procedure analysis (Q2), mitigation strategies generation (Q3), and patch code generation (Q5). 
Q4 is a supplement to Q1. After learning the procedure of attacks, we aim to extract elements related to vulnerabilities from the dependency graph, which do not exist within the original textual descriptions.

Furthermore, in each step, we integrated structural text information and static analysis results into prompts to reduce LLM's randomness. Specifically, we utilized vulnerability description and CDG in Q2. Q3 embeds attack procedures and CDG. Q4 is queried by combining attack procedures, CDG, and generated strategies. Afterwards, we supplemented CDG as well as the program slices. Eventually, we integrated CDG, program slices, and mitigation strategies into prompt of Q5 to generate Vul-Fix Pairs.

\begin{figure}
    \centering
    \includegraphics[width=\linewidth]{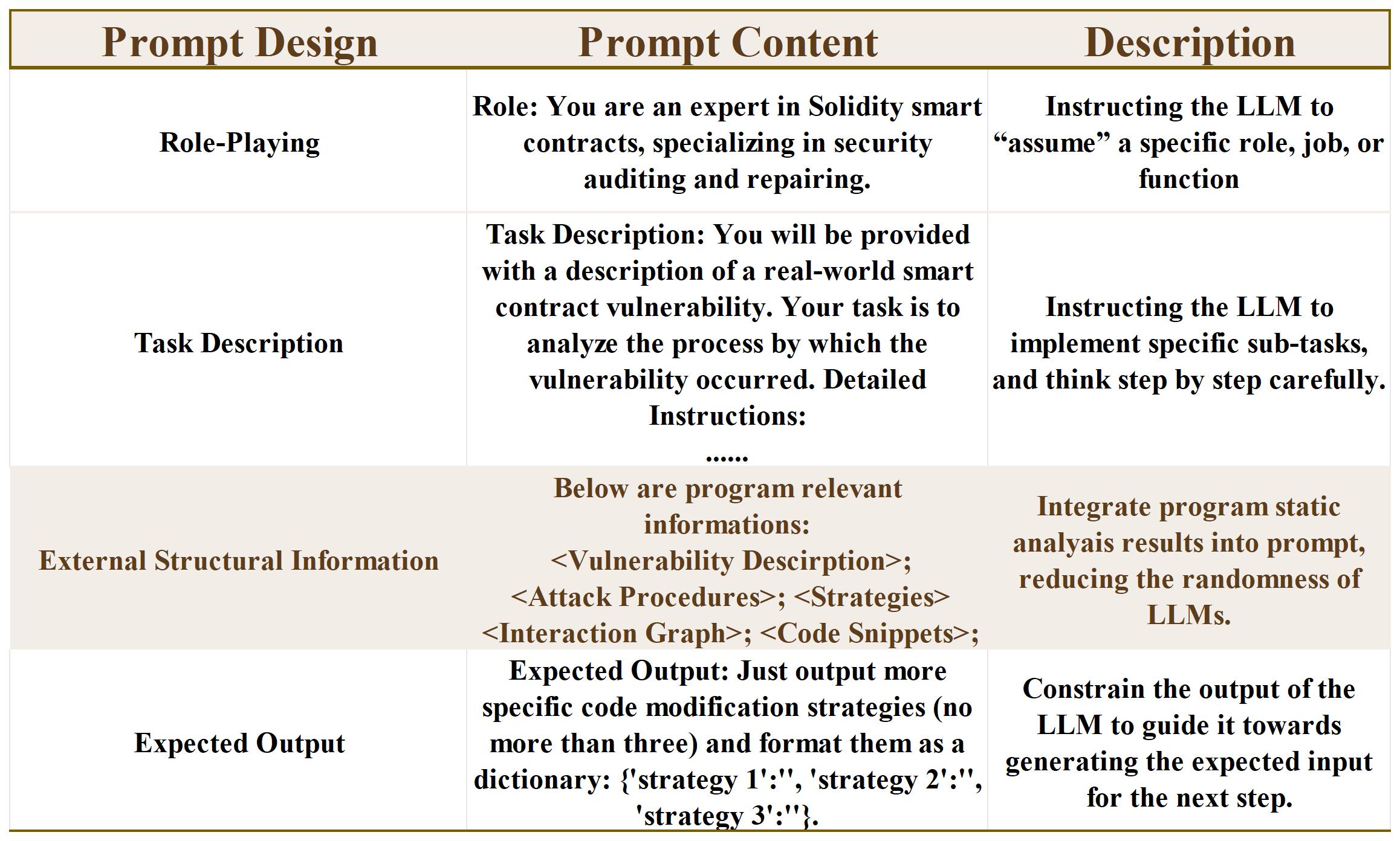}
    \vspace{-2em}
    \caption{Prompt Design of \repair}
    \label{fig:prompt}
\vspace{-2em}
\end{figure}

For all prompts, we designed a general template as shown in Figure~\ref{fig:prompt}. Its content mainly consists of several core parts: Role-Playing, Task Description, Expected Output, and External Information. We have open-sourced all prompt templates in the GitHub repository.

\subsection{\textbf{Patch Refinement}}
In this section, we introduce patch refinement which comprises of LLM Validator, Refinement, and Manual Checker. In particular, we adopt the concept of multi-agent debate~\cite{debate} that another LLM is leveraged to judge whether the patches fixed the vulnerable code. If not, then the fixed code is inputted into Q6 in Figure~\ref{fig:framework} combined with supplemental recommendations. Finally, the fixed code is manually evaluated.

\section{The \repair Tool}

\subsection{Implementation}
The \repair consists of the following core modules: 
\setlength{\leftmargini}{0.6cm}
\begin{itemize}
    \item \textit{Patch Generator} is the main module for generating patches. It implements vulnerability localization and constructs a pipeline for the entire generation task.
    \item \textit{Document Parser} mainly focuses on parsing input audit reports, and it can be customized to support other types of audit formats.
    \item \textit{Program Analyzer} is used to analyze the entire project, and construct dependency or data flow graphs to enhance the effectiveness of LLM-based CoT.
    \item \textit{LLM Interface} is designed with several interfaces, enabling developers to utilize different LLMs with customized configurations.
    \item \textit{Prompt utils} provides different prompt templates adopted for our sub-tasks. These prompt templates can be customized for specific sub-task settings.
\end{itemize}
Furthermore, in this paper, we utilized Slither\cite{slither} as primary program static analysis tool. We implemented a flexible interface to employ different LLMs, such as Llama, GPT-4, GPT-3.5, CodeT5. We used GPT-4 and GPT-3.5 as base model in our experiment.

\begin{figure}
    \centering
    \includegraphics[width=\linewidth]{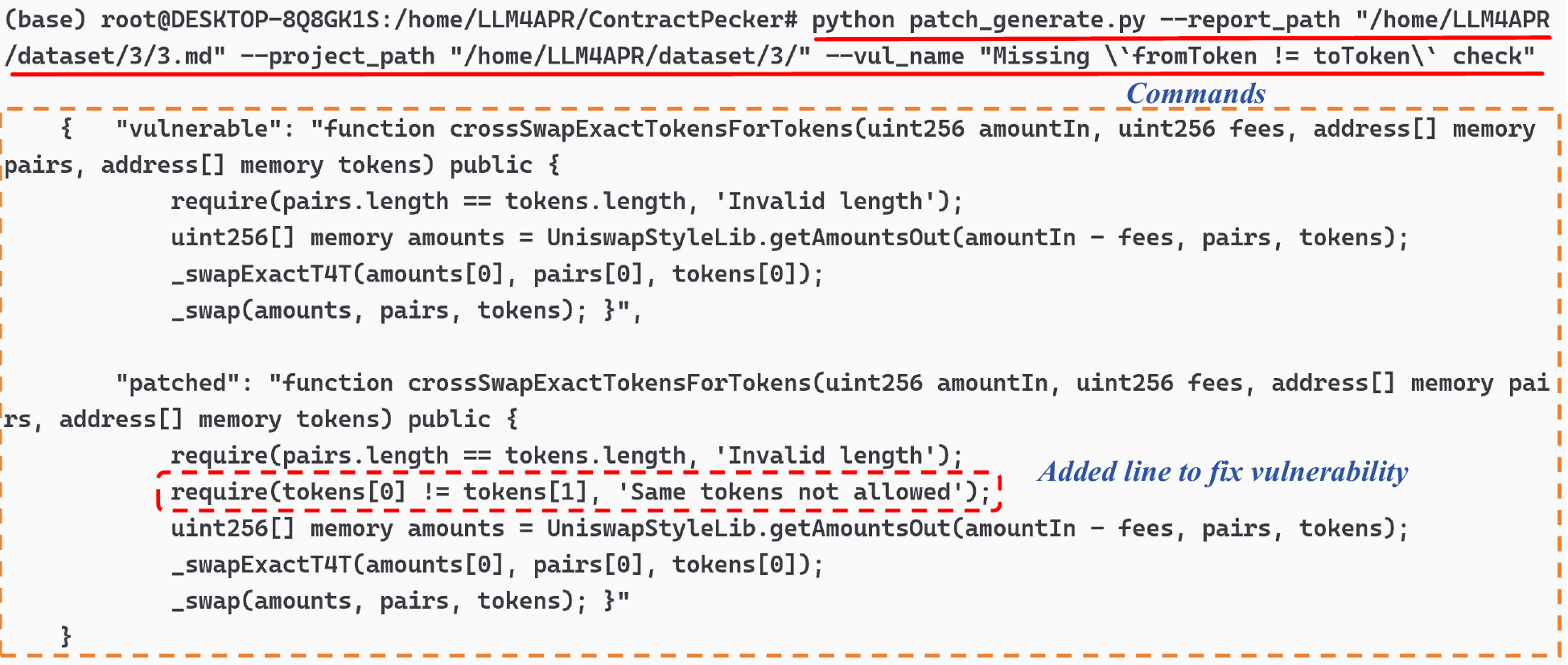}
\vspace{-2em}
    \caption{The Screenshot of \repair}
\label{fig:tool_screen}
    \vspace{-2.5em}
\end{figure}

\vspace{-0.5em}
\subsection{Usage}
We provide a command-line interface for developers to use our tool \repair. The usage is simple, as presented in Figure~\ref{fig:tool_screen}. The detail of arguments is shown below to illustrate its usage:
\setlength{\leftmargini}{0.6cm}
\begin{itemize}
    \item \textit{---report\_path} and \textit{---project\_path} are used to specify the exact report and project paths of the corresponding vulnerabilities.
    \item \textit{---vul\_name} specifies the vulnerability that needs to be repaired as included in the audit report.
    \item \textit{---solc\ remaps} is also available for associating particular external dependency libraries (e.g., OpenZeppelin) to this project.
    \item \textit{---output\ path} is used to store generated patches.
\end{itemize}
After preparing all inputs, the user can use the \textit{CompilationChecker} module to test if the project can compile under the current environment. If the project cannot compile, the user needs to download the required packages as suggested by \textit{CompilationChecker}. Then, the user runs \textit{Python} \textit{patch\_generate.py} with the aforementioned arguments to generate patches for the specified vulnerability. Finally, the patch will be stored in the designated path. The orange box in Figure~\ref{fig:tool_screen} illustrates an example of a vulnerability and its corresponding patch.

\section{Preliminary Evaluation}
\subsection{Data Collection and Metrics}
In this experiment, we collected 48 high-risk findings that contain both vulnerabilities and their fix recommendations from Code4Rena platform. The main metric to evaluate patches generated by the \repair is accuracy. Additionally, we added strategy success rate, which is inspired by the Hit Rate (HR)\cite{hitrate} from recommendation systems, to evaluate the quality of generated strategies.

\subsection{Evaluations}
\subsubsection{\textbf{Strategies Comparison}}
Before generating code patches, it is essential to demonstrate that \repair can indeed provide correct mitigation strategies.
We conduct an experiment, that uses the intermediate result to determine whether the fix recommendation matches one of the mitigation strategies generated by the \repair. We adopt the strategy success rate metric to evaluate its effectiveness. 
Success Rate indicates the rate of single mitigation strategy generated by \repair matches the fix recommendation.
Top-3 Success Rate measures whether one of the top three mitigation strategies generated by \repair matches the fix recommendation extracted from the audit report. 

\obs{{\textit{\textbf{Results 1:} According to our evaluations, the \textbf{Success Rate} of \repair with GPT-3.5 in generating strategy is \textbf{70.4\%}, and the \textbf{Top3 Success Rate} in generating mitigation strategies is \textbf{91.6\%}.}}}

\subsubsection{\textbf{Patches Evaluation}}
In this section, we aim to evaluate the quality of the generated patches. Due to the lack of a finely labeled dataset, we employ three mechanisms to assess the accuracy of the patches: GPT-4 Check, Manual Check, and Compilation Check. Specifically, in compilation check, we compile the contract to examine if it can compile after incorporating the patches. In Validator check and manual check, we leverage GPT-4 and manual efforts to validate the functional correctness of the generated patches, respectively. Additionally, we categorize all patches into four classes: 1. Valid Patches that pass compilation, 2. Valid Patches that need minor modification, 3. Patches that need logic modification, and 4. Irreparable Patches.

Valid Patches that pass compilation represent patches that have passed all three check mechanisms. Valid Patches that need minor modification are those that have passed both the Validator Check and Manual Check but failed the Compilation Check. These patches contain minor issues, such as only included fixed code snippets directly related to the vulnerability functions or contained pseudo variable names that need to be replaced with state variable names defined in the contract. Thus, these patches require developers to make minor modifications without any logic changes.

Patches that need logic modification denote patches that have only partially implemented the necessary code. Since we input only relevant code snippets into our tool, it cannot fully understand the entire business logic of the projects. Therefore, the core logic code related to business functionalities is too complex for \repair to implement correctly. Such as how to define time-weighted average price (TWAP) in a function when facing price manipulate attacks.

Irreparable Patches refer to patches where the vulnerability cannot be fixed by our tool. This can occur for several reasons: 1. The vulnerability description is too vague for \repair to locate the bug and understand how the attack happens. 2. The business logic of the vulnerability code is too complex, leading \repair to produce significant hallucinations. 3. Even if \repair understands how the attack happens, it might not extract the correct target functions directly related to the vulnerability from description.

\obs{\textit{\textbf{Results 2:} Based on our experiment, \repair can generate \textbf{48\%} of valid patches that \textbf{fix the vulnerabilities}. \textbf{21\%} of patches require \textbf{only minor modifications}.}}

\section{RELATED WORK}
The repair of smart contracts has been widely discussed over the past few years, while few of them can address real-world vulnerabilities. SGuard\cite{sguard} relies on pre-defined patterns to fix four low-level vulnerabilities. Rodler \cite{evmpatch} proposed EVMPatch, focusing on bytecode, and also only fixes low-level vulnerabilities. SCRepair\cite{screpair} is a search based method limited by manually created unit tests. 
Smartfix\cite{smartfix} is the most recent proposed tool which uses statistical models to guide repair, but it is also limited to predefined vulnerability fixes. To our knowledge, these methods still focus on commonly known vulnerabilities and do not address real-world vulnerabilities that require comprehension of business logic or high-level semantics.

\vspace{-0.5em}
\section{CONCLUSIONS}
In this paper, we proposed \repair, an LLM-empowered repair tool for real-world smart contract projects. We utilized the Chain-Of-Thought concept that breaks down generation task into sub-tasks, in which we integrated program static analysis, including dependency analysis and program slicing. This design enhanced the accuracy of the LLM by combining high-level semantic and static analysis. Our preliminary evaluation demonstrates the effectiveness of \repair.

\vspace{-0.5em}
\begin{acks}
This work is supported by 
National Natural Science Foundation of China (62202011, 62172010).
\end{acks}

\bibliographystyle{ACM-Reference-Format}
\bibliography{sample-base}

\end{document}